# Full Geometric Control of Hidden Color Information in Diffraction Gratings under Angled White Light Illumination

**Authors:** John You En Chan [1], Qifeng Ruan [1,2] *, Hongtao Wang [1,3], Hao Wang [1], Hailong Liu [4], Zhiyuan Yan [3], Cheng-Wei Qiu [3], Joel K.W. Yang [1,4] *

[1] Engineering Product Development, Singapore University of Technology and Design, Singapore 487372.

[2] Ministry of Industry and Information Technology Key Lab of Micro-Nano Optoelectronic Information System, Harbin Institute of Technology (Shenzhen), Shenzhen 518055, P. R. China.

[3] Department of Electrical and Computer Engineering, National University of Singapore, Singapore 117583.

[4] Institute of Materials Research and Engineering, A*STAR (Agency for Science, Technology and Research), Singapore 138634.

* To whom correspondence should be addressed. Email: ruanqifeng@hit.edu.cn; joel_yang@sutd.edu.sg

# ABSTRACT

Under white light illumination, gratings produce an angular distribution of wavelengths dependent on the diffraction order and geometric parameters. However, previous studies of gratings are limited to at least one geometric parameter (height, periodicity, orientation, angle of incidence) kept constant. Here, we vary all geometric parameters in the gratings using a versatile nanofabrication technique, two-photon polymerization lithography, to encode hidden color information through 2 design approaches. The first approach hides color information by decoupling the effects of grating height and periodicity under normal and oblique incidence. The second approach hides multiple sets of color information by arranging gratings in sectors around semi-circular pixels. Different images are revealed with negligible crosstalk under oblique incidence and varying sample rotation angles. Our analysis shows that an angular separation ≥ 10° between adjacent sectors is required to suppress crosstalk. This work has potential applications in information storage and security watermarks.

# INTRODUCTION

Diffraction gratings are ubiquitous in optical systems with established applications in displays[1-4], filters[5-9], sensors[10-13], and optical security devices[14-20]. Grating-based optical security devices are of particular interest as they offer animated[21], three-dimensional[22], and rainbow effects[23, 24] that are easy to be verified and difficult to be copied. Colors in gratings are produced by tuning geometric parameters[25-27] such as the height, periodicity, orientation, and angle of incidence. However, the combined use of these parameters to hide color information in gratings has not been achieved efficiently. Though previous works have used polarization selectivity[28, 29] and directional stretching[30] to encode color information in gratings, they are limited to only 2 independent polarization states or stretching directions that reveal the encoded information with no crosstalk. Hence, a solution to increase the information storage capacity of gratings is necessary. Here, we show up to 5 sets (potentially 18 sets) of hidden color information in gratings entirely by strategic combinations of their geometric parameters using two design approaches. The first approach exploits the decoupling effect between grating height and periodicity, so that an image appears polychromatic under normal incidence, but monochromatic under oblique incidence, and vice versa. The second approach exploits the directional effect of angularly multiplexed gratings arranged in sectors around semi-circular pixels to encode multiple sets of hidden color information. Instead of full circles, semi-circular pixels are chosen as they can be fabricated more efficiently to achieve almost the same desired optical effect because of rotational symmetry possessed by the gratings. When these gratings are illuminated under oblique incidence, diffracted light from each sector propagates in different directions. If the angular separation between adjacent sectors is larger than a minimum value (10° based on our analysis), then only the direction of diffracted light from one sector is collected by the microscope objective at a specific rotation angle. By rotating the gratings in steps of the angular separation, each set of hidden color information is revealed with negligible crosstalk. The gratings in our work only require illumination by one white light source, which is simpler compared to previous works[1-3] that require additional light sources and waveguides for displaying color information. Furthermore, we use two-photon polymerization lithography (TPL) to fabricate the gratings in IP-Dip photoresist on a fused silica substrate (see Methods). TPL is a nanofabrication technique whereby a high-intensity femtosecond laser polymerizes a localized volume in the photoresist through nonlinear absorption[31]. By scanning the laser through the photoresist, complex 3D structures with sub-micron resolution are fabricated[32-42]. As this technique enables great design flexibility and geometric control of the structures, it is suitable for fabricating gratings with multiple sets of hidden color information that have potential applications in information storage and security watermarks[43, 44].

## RESULTS

### Design concepts

The setup for observing colors produced from an array of diffraction gratings (referred to as the sample) is illustrated in Fig. 1a. White light from an optical fiber illuminates the sample, and the transmitted light collected by the microscope objective forms a color image in camera. We denote $\theta_i$ as the polar angle, and $\phi_i$ as the azimuthal angle of incident light respectively. Throughout this paper, we define $\phi_i = 0°$ (not shown in figures). Multiple color images are encoded in angularly multiplexed gratings arranged in sectors around semi-circular pixels. Under oblique incidence, different images are individually revealed by rotating the sample in clockwise direction to angles $R$ corresponding to the orientation of each sector (Fig. 1b). Each rotation step equals the angular separation between adjacent sectors. However, when the sample is rotated to an angle in between adjacent sectors, a faint image showing crosstalk is observed (Fig. S1). This concept of revealing hidden color information can be understood by the diffraction effect that each grating sector imposes on obliquely incident light (Fig. 1c). Each sector is characterized by its geometric parameters: height $H$, width $W$, and periodicity $P$. The orientation of each sector is represented by a grating vector pointing in the direction of periodicity. The plane of incidence (POI) contains the incident light ray and surface normal, but it does not necessarily contain the grating vector and $m^{th}$ order diffracted light. Here, we denote $\sigma$ as the azimuthal angle between the POI and grating vector; $\theta_m$ is the polar angle, and $\phi_m$ is the azimuthal angle of the $m^{th}$ order diffracted light relative to the POI. The optical behavior of transmission gratings in directional cosine space[45] is expressed in Equation 1 (see Supporting Information for derivation):

$$\alpha_m - \alpha_i = -\frac{m\lambda}{P}\cos\sigma, \beta_m - \beta_i = -\frac{m\lambda}{P}\sin\sigma \qquad (1)$$

where $m$ is the diffraction order, $\lambda$ is the wavelength of light, $\alpha_{i(m)}$ and $\beta_{i(m)}$ are the directional cosines for incident ($m^{th}$ order diffracted) light respectively. The directional cosines are further expressed in Equation 2:

$$\alpha_{i(m)} = \sin\theta_{i(m)}\cos\phi_{i(m)}, \beta_{i(m)} = \sin\theta_{i(m)}\sin\phi_{i(m)} \qquad (2)$$

We only consider wavelengths of diffracted light collected by the microscope objective as they contribute to the colors of the observed image. This condition is expressed in Equation 3:

$$|\sin\theta_m| \leq NA \qquad (3)$$

where $NA$ refers to the numerical aperture of the microscope objective. Hence, these equations provide the theoretical basis for designing gratings with hidden color information.

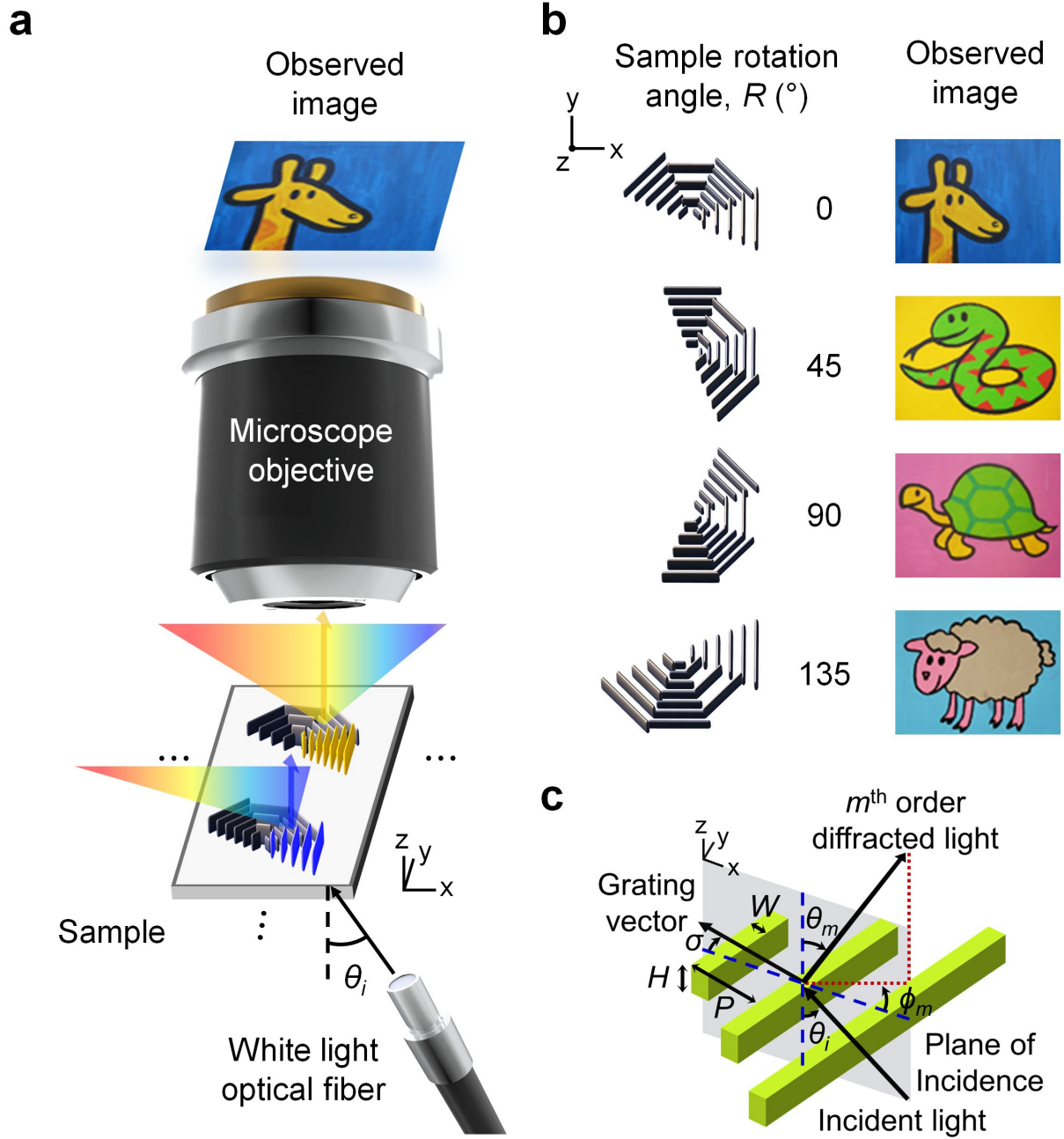

**Fig. 1. Design schematics.** (**a**) Setup for observing the sample that comprises an array of semi-circular pixels with angularly multiplexed gratings arranged in sectors around each pixel. These gratings encode different sets of color information. When white light illuminates the sample at a specific angle, the light collected by the microscope objective forms a colorful observed image. The incident light is parameterized by its polar angle $\theta_i$, whereas its azimuthal angle $\phi_i$ (not shown) is defined to be 0°. (**b**) Under oblique incidence, different images are observed at sample rotation angles $R$ in the clockwise direction. The sample is rotated in steps of the angular separation between adjacent sectors. (**c**) Diffraction of light in a grating with height $H$, width $W$, and periodicity $P$. $\sigma$ is the azimuthal angle between the grating vector and plane of incidence (POI). The POI contains the incident light ray and surface normal, but it does not necessarily contain the grating vector and $m^{th}$ order diffracted light. $\theta_m$ ($\phi_m$) is the polar (azimuthal) angle of the $m^{th}$ order diffracted light relative to the POI.

## Colors produced by gratings in experiments

We used two-photon polymerization lithography (TPL) to directly fabricate gratings with $H = 0.6 - 1.8$ µm and $P = 0.8 - 1.6$ µm. $W$ (~ 0.3 µm) was fixed by the smallest printable and mechanically stable feature size of our TPL machine. The gratings were made of UV-cured IP-Dip photoresist (average $n_g \approx 1.55$, Fig. S2) on a fused silica substrate. When the gratings were illuminated under normal ($\theta_i = 0°$) and oblique ($\theta_i = 45°$) incidence, we observed two different color palettes using an optical microscope with $NA = 0.1$ (Figs. 2a – b). The observed images under $\theta_i = 0°$ ($\theta_i = 45°$) appeared to have a light (dark) background because directly transmitted light was collected (not collected) by the microscope objective. We arbitrarily selected a grating ($H = 0.8$ µm, $P = 1.0$ µm) for spectral analysis (Fig. 2c). Under $\theta_i = 0°$, a transmittance peak was formed at $\lambda = 430$ nm, whereas under $\theta_i = 45°$, the peak red-shifted to $\lambda = 707$ nm with reduced intensity and narrower bandwidth. These peak positions can be attributed to two different mechanisms that operate under $\theta_i = 0°$ and $\theta_i = 45°$ respectively to produce color. Under $\theta_i = 0°$, colors are caused by the interference effect of light between the grating and its surrounding environment[30]. The peak wavelength of the zeroth order light ($m = 0$) is estimated in Equation 4 (see Supporting Information for derivation):

$$\lambda_0 = \frac{H(n_g - n_e)}{q} \tag{4}$$

where $q$ is a positive integer, $n_g$ is the refractive index of the grating structure, and $n_e$ is the refractive index of the environment (in this case, air). Under $\theta_i = 45°$, colors are caused by the wavelength-dependent angles of diffracted light collected by the microscope objective. The peak wavelength of the $m^{th}$ diffraction order ($m \neq 0$) light collected by the microscope objective is expressed in Equation 5:

$$\lambda_m = \frac{P \sin \theta_i}{m} \tag{5}$$

This peak is caused by diffracted light that enters the microscope objective normally along the optical axis ($\theta_m = 0°$), where the intensity of collected light is maximum compared to other directions. Maximum intensity is achieved in the case of planar diffraction, in which the grating vector is parallel to the plane of incidence ($\sigma = 0°$). Equations 4 – 5 suggest that $H$ and $P$ can be varied independently to produce different colors under normal and oblique incidence. As shown in Figs. 2a – b, variations in $H$ produced greater color change under $\theta_i = 0°$, whereas variations in $P$ produced greater color change under $\theta_i = 45°$. Moreover, the color palette under $\theta_i = 45°$ appeared more saturated than the color palette under $\theta_i = 0°$. This observation was verified in the CIE diagram (Fig. S3), which showed that the chromaticity coordinates for $\theta_i = 45°$ were spread out towards the periphery, whereas the chromaticity coordinates for $\theta_i = 0°$ were clustered around the center. This saturation difference between both palettes can be explained by the range of wavelengths collected by the microscope objective. Under $\theta_i = 0°$, all wavelengths for the zeroth order ($m = 0$) are collected. However, under $\theta_i = 45°$, only limited ranges of wavelengths for non-zero orders ($m \neq 0$) are collected. The range of collected wavelengths for the $m^{th}$ diffraction order is expressed in Equation 6 (see Supporting Information for derivation):

$$\Delta \lambda_m = \frac{2P(NA)}{m} \tag{6}$$

By using a small $NA$ value, a narrow band of wavelengths will be collected by the microscope objective and produce high color saturation. In this work, we used a microscope objective with $NA = 0.1$, yielding $\Delta \lambda_m = 200$ nm for $P = 1.0$ µm and $m = 1$.

Under $\theta_i = 45°$, we also observed that variations in $H$ caused significant changes in color brightness for gratings with the same $P$ (Fig. 2b). The transmission spectra of these gratings showed almost constant peak wavelength positions but varying peak intensities (Fig. S4). By using scalar diffraction theory[25, 26], the intensity distribution of the $m^{th}$ diffraction order ($m \neq 0$) can be approximated from Equation 7:

$$\eta_m = \frac{4}{\pi^2 m^2} \sin^2(\pi m D) \sin^2\left(\frac{\Delta \rho}{2}\right) \tag{7}$$

where $D = (W / P)$ is the grating duty cycle, and $\Delta \rho = 2\pi H / \lambda_m$ is the phase modulation of light induced by the gratings. Though scalar diffraction theory is only expected to be accurate for coarse and shallow gratings (i.e. $P / \lambda \geq 10$ and $H / \lambda \leq 1$)[46-49], Equation 7 still showed reasonable agreement with the observed trend in experimental data. Separately, the color brightness under $\theta_i = 45°$ can also be controlled by varying the angle $\sigma$ between the grating vector and plane of incidence. To demonstrate the change in color brightness, we fabricated and observed another set of gratings with $H$ = 0.8 μm, $P$ = 1.0 μm and $\sigma$ = 0° – 10° (Fig. 2d). The gratings appeared darker as $\sigma$ increased, and the grating with $\sigma$ = 10° was barely visible compared to $\sigma$ = 0°. The transmittance spectra of the gratings showed that the peak wavelength position remained unchanged, but the peak intensity decreased rapidly from $\sigma$ = 0° to 4°. This decrease in intensity can be explained using directional cosine diagrams. For clarity, we show only the spectra for diffraction order $m$ = 1 in each diagram of Fig. 2e, whereas the diagrams for $m$ = 0 and $m$ = 2 are shown in Fig. S5. The microscope objective has $NA$ = 0.1, which is represented by a dashed circular region around the origin ($\alpha$ = 0°, $\beta$ = 0°). At $\sigma$ = 0°, the diffracted wavelengths produce a horizontal spectra line that passes through the origin. Only a limited range of the diffracted wavelengths is contained within the $NA$ region. At $\sigma$ = 4°, the spectra line is rotated anti-clockwise by the same angle, but still partially contained within the $NA$ region. This rotation suggests a loss in intensity, which was experimentally verified in Fig. 2d. At $\sigma$ = 10°, the spectra line had completely exceeded the $NA$ region, resulting in effectively zero intensity. Hence, to avoid crosstalk, the minimum angular separation between grating sectors is the angle $\sigma$ at which the spectra line just exceeds the $NA$ region in the directional cosine diagram. This minimum value assumes that the gratings comprise an infinite number of periodic lines. However, fabricated gratings have a limited number of lines, which can weaken the directionality of the grating vectors and result in crosstalk. In practice, the angular separation between grating sectors should be designed larger than $\sigma$ = 10° to suppress crosstalk.

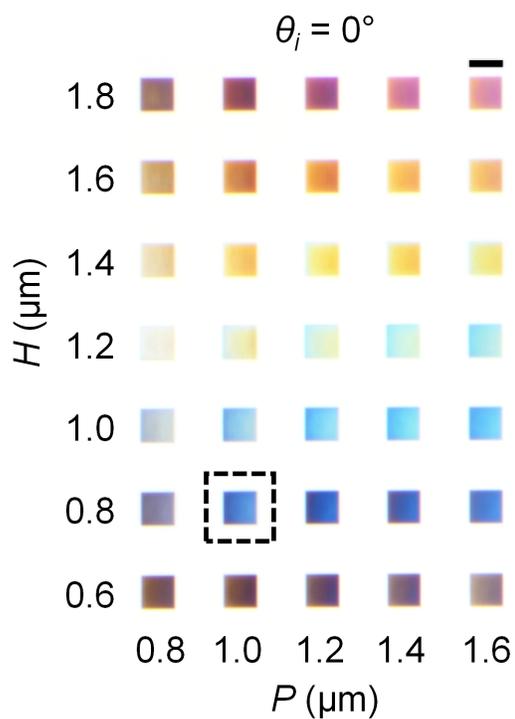
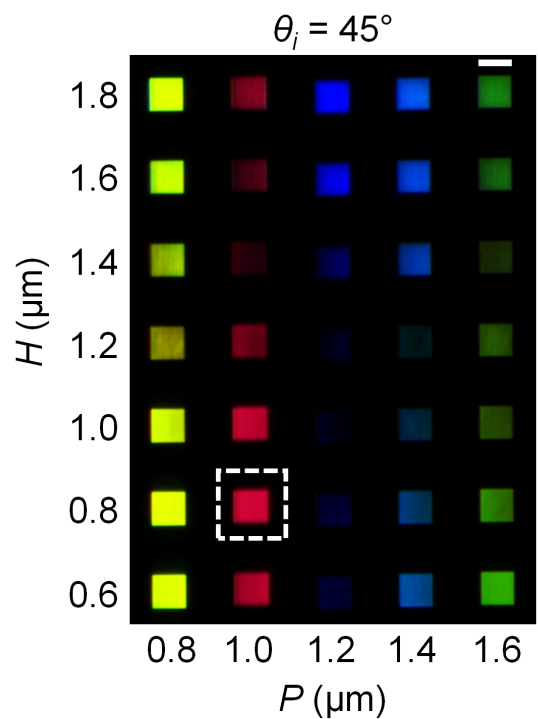
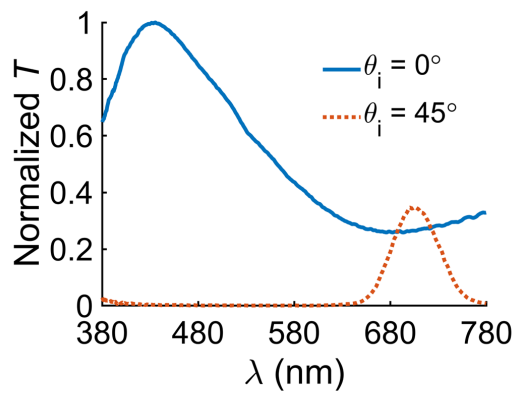
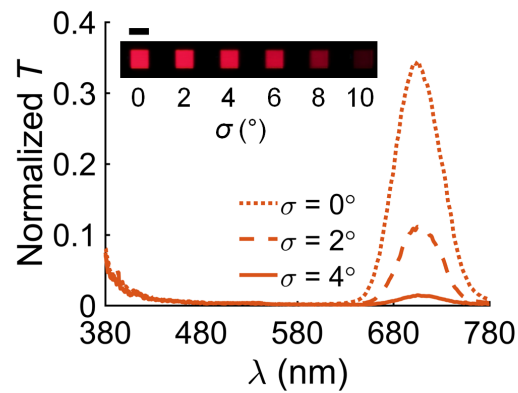
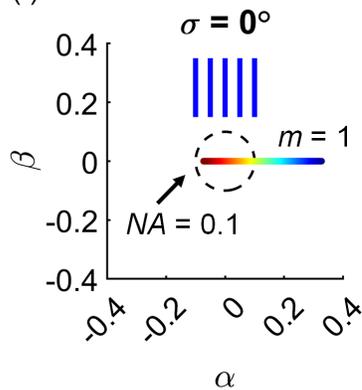
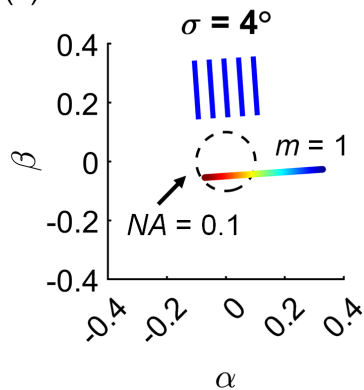
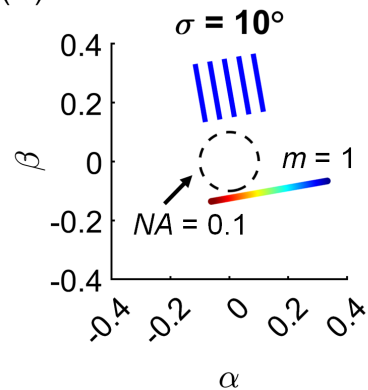

**Fig. 2. Analysis of experimentally measured color palette.** (**a–b**) Transmission optical micrographs (scale bar: 60 μm) of the color palette under (**a**) normal incidence $\theta_i = 0°$ and (**b**) oblique incidence $\theta_i = 45°$, for gratings with varying height $H$ and periodicity $P$. A grating ($H = 0.8$ μm and $P = 1.0$ μm), indicated by the dashed outline of a square, is arbitrarily selected for spectral analysis. (**c**) Normalized transmittance spectra $T$ of the selected grating under $\theta_i = 0°$ and $\theta_i = 45°$. (**d**) Normalized transmittance spectra of the selected grating for $\sigma = [0, 2, 4]°$ under $\theta_i = 45°$. $\sigma$ is the azimuthal angle between the grating vector and plane of incidence. Scale bar: 60 μm. (**e**) Directional cosine diagrams for the selected grating with (i) $\sigma = 0°$, (ii) $\sigma = 4°$, (iii) $\sigma = 10°$. These diagrams are plotted for $\theta_i = 45°$, wavelengths $\lambda = 380 - 780$ nm (blue to red), and diffraction order $m = 1$. The dashed circle represents numerical aperture $NA = 0.1$.

## Hidden color images in gratings

Here, we demonstrate the two design approaches to encode hidden color information in gratings. Our first approach exploits the decoupling effect of grating height and periodicity under normal and oblique incidences. For this approach, we printed an image[50] of "elephant inside snake" using regular gratings with the same orientation and periodicity ($P = 1.2$ μm), but varying heights ($H = 1.4 - 1.8$ μm). Under normal incidence $\theta_i = 0°$, the elephant and snake were easily identified in the image due to their distinct color contrast (Fig. 3a(i)). However, under oblique incidence $\theta_i = 45°$, the same image appeared like a blue hat (Fig. 3a(ii)). The image also appeared to consist of square patches with non-uniform colors. The square patches were caused by stitching errors between patterning areas, whereas non-uniform colors were caused by machine errors in finding the interface between the photoresist and substrate. If necessary, these errors can be mitigated by using adaptive stitching algorithms[51] and printing an additional base layer beneath the gratings[30]. A magnified scanning electron microscope (SEM) image of the gratings with different heights that comprised the "elephant inside snake" is shown in Fig. 3a(iii).

Our second approach exploits the directional effect of angularly multiplexed gratings arranged in sectors around a semi-circular pixel. Four different animal paintings (giraffe, snake, turtle, sheep)[52-55] were encoded into four grating sectors with varying heights ($H = 0.8 - 2.0$ μm), periodicities ($P = 0.8 - 1.6$ μm) and orientations in each pixel. As the sample was rotated to angles $R = [0, 45, 90, 135]°$, the individual paintings were revealed with negligible crosstalk (Figs. 3b(i – iv)) because their angular separation (45°) was much larger than the minimum limit (10°). However, small amounts of crosstalk can still be discerned in Fig. 3b(iv) due to the small pixel size with only a few repeating lines per grating sector. We found that grating sectors with 3 lines or more, as shown in the high-magnification SEM image of a constituent pixel (Fig. 3b(vi)), were sufficient to produce acceptable color pixels. Fabricating larger pixels with more lines per grating sector can help to suppress crosstalk, but also lead to longer fabrication time and a more pixelated image appearance. To visualize how the individual paintings are revealed by sample rotation, the directional cosine diagrams for one pixel are shown in Fig. S6. Under $\theta_i = 0°$, the paintings could not be distinctly identified, instead we observed a noisy image that appeared like a superposition of the paintings (Fig. S7a). This superposition can be explained by the collection of all zero-order ($m = 0$) light from the gratings. The superposed appearance was also observed in the low-magnification SEM image (Fig. 3b(v)).

Finally, we demonstrate that both approaches can be combined, and the intensity of pixels switched on and off arbitrarily to reveal specific shapes within the same image. A famous example is Rubin's vase[56], which is an ambiguous image that can be perceived as a vase or two faces due to the figure-ground distinction. To unravel this ambiguous image, we designed each shape with different angularly multiplexed gratings. All shapes have the same grating height ($H = 0.8$ μm), but each shape has a different grating periodicity $P = 0.8 - 1.2$ μm. In each shape, each pixel was designed with 5 grating sectors (angular separation = 36°), but the sectors were selectively printed or omitted, so that the pixel would appear bright or dark at the corresponding angle of sample rotation. The pixels that comprised each shape of Rubin's vase are shown in the SEM images of Fig. 4b. Under $\theta_i = 45°$, the sample was rotated to $R = [0, 36, 72, 108, 144]°$, which selectively revealed the different colored shapes of Rubin's vase (Fig. 4a). Under $\theta_i = 0°$, Rubin's vase appeared monochromatic, in which darker (lighter) shapes represent a higher (lower) fill factor of the gratings (Fig. S7b). The fill factor $F$ can be estimated by the grating area per unit cell, which is expressed in Equation 8:

$$F = \frac{0.5(N_L P)^2 \sin A}{S^2} \times N_S \frac{W}{P} \times 100\% \tag{8}$$

where $S$ = 12 µm is the center-to-center separation between adjacent pixels; $N_S$ = [3, 2, 3] is the number of printed sectors in a pixel; $N_L$ = [6, 3, 5] is the number of lines per sector; $A$ = 36° is the angular span of each sector; $W$ = 0.3 µm is the grating width; $P$ = [1.0, 0.8, 1.2] µm is the grating periodicity. We estimated $F$ to be [6.6, 0.9, 5.5] % for the left face, center vase, and right face of the image respectively. Apart from revealing Rubin's vase, we also used this combined approach to design gratings encoded with multiple hidden quick response codes in one sample (Fig. S8).

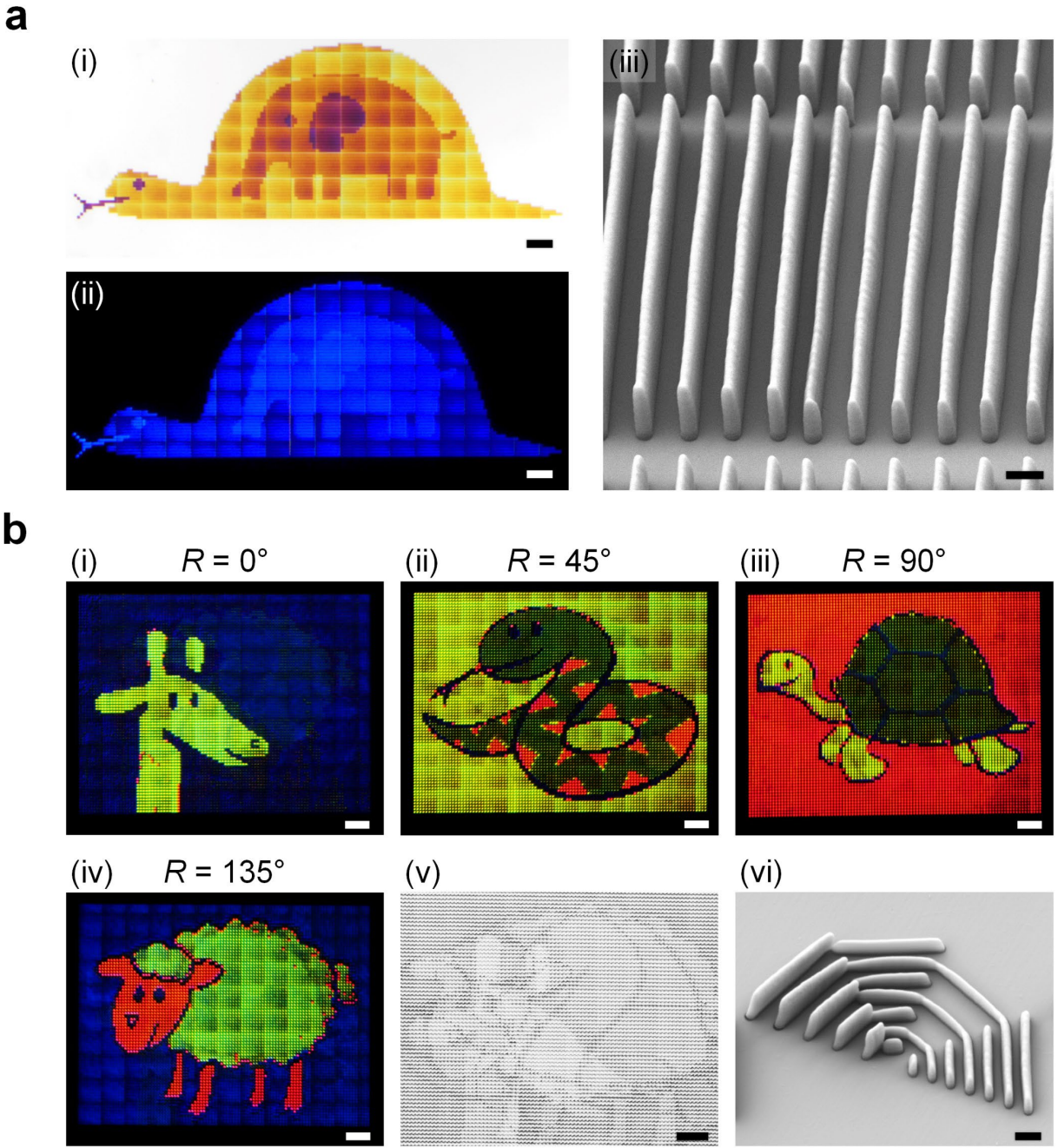

**Fig. 3. Hidden color information revealed through oblique incidence and varying sample rotation angles.** (**a**) (i – ii) Optical micrographs of "elephant inside snake" for (i) normal incidence $\theta_i = 0°$, and (ii) oblique incidence $\theta_i = 45°$. Scale bar: 100 µm. (iii) High-magnification SEM image of gratings with the same periodicity ($P = 1.2$ µm) but different heights ($H = 1.4 – 1.8$ µm) in "elephant inside snake". Scale bar: 1 µm. (**b**) (i – iv) Optical micrographs of the individual animal paintings under $\theta_i = 45°$ and sample rotation angles $R = [0, 45, 90, 135]°$ respectively. Scale bar: 100 µm. (v) Low-magnification SEM image of the sample with encoded paintings. Scale bar: 100 µm. (vi) High-magnification SEM image of a pixel with angularly multiplexed gratings used to encode the paintings. Scale bar: 1 µm.

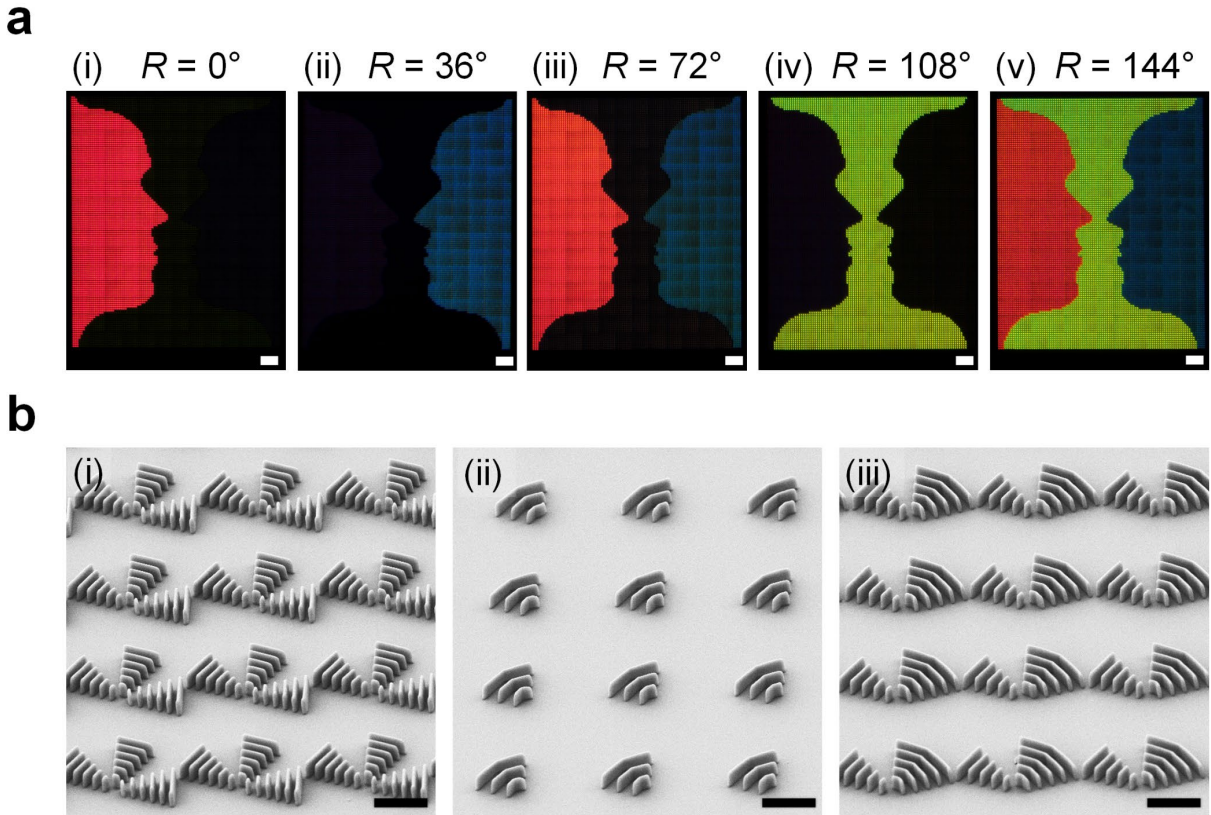

**Fig. 4. Revealing Rubin's vase.** (**a**) Optical micrographs of Rubin's vase under oblique incidence ($\theta_i$ = 45°) and sample rotation angles $R$ = [0, 36, 72, 108, 144]°. Scale bar: 100 μm. (**b**) SEM images of pixels in the (i) left face, (ii) center vase, (iii) right face of Rubin's vase. Scale bar: 5 μm.

## DISCUSSION

We have examined how the combined variation of geometric parameters (height, periodicity, orientation, angle of incidence) was used to encode hidden color information in diffraction gratings through two design approaches. The first approach exploited the decoupling effect of grating height and periodicity to design images that appear polychromatic under normal incidence ($\theta_i$ = 0°), but monochromatic under oblique incidence ($\theta_i$ = 45°), and vice versa. The second approach exploited the directional effect of angularly multiplexed gratings arranged in sectors around semi-circular pixels to design up to 5 sets of hidden color information. These sets of information were individually revealed with negligible crosstalk under oblique incidence ($\theta_i$ = 45°) by rotating the sample to specific angles in steps of the angular separation between adjacent sectors. The minimum angular separation (10°) to suppress crosstalk was determined graphically by the angle at which the spectra line for a diffraction order ($m \neq 0$) just exceeds the *NA* region in the directional cosine diagram. Hence, potentially up to 18 sets of color information can be hidden in angular multiplexed gratings. To reveal the hidden sets of color information, only broadband light from an optical fiber bundle is required to illuminate the gratings. We used TPL to fabricate the gratings because it enables great design flexibility and control of geometric parameters. The fabrication took around 2 hours for an area of 1 mm$^2$. To improve the throughput, parallel processing TPL systems can be used[57-59]. After fabrication, the gratings can be replicated by nanoimprinting techniques such as molding for mass production[16]. These gratings are mechanically stable under room temperature and pressure, and they can be packaged in transparent casings for protection. We anticipate that gratings with hidden color information will find potential applications in information storage and security watermarks.

## METHODS

### Fabrication

A drop of IP-Dip photoresist was placed on a fused silica substrate, which was then transferred to a Nanoscribe GmbH Photonic Professional GT system for patterning the gratings using a laser power of 10 mW and laser scan speed of 500 µm/s. The height of the gratings was controlled by patterning each additional layer with a discrete height step of 0.2 µm. After patterning, the substrate was removed from the system and developed in chemical solution. During development, the substrate was immersed in propylene glycol monomethyl ether acetate solution for 5 min, then in isopropyl alcohol solution with UV curing (Dymax BlueWave MX-150 set to 60% maximum power) for 5 min, and then in nonafluorobutyl methyl ether solution for 5 min. Lastly, the substrate was dried in air.

### Optical imaging and color measurements

The optical micrographs in Figs. 2 – 4 were captured with a Thorlabs TL2X-SAP NA0.10 objective and a Nikon DS-Ri2 camera. A Thorlabs OSL2 fibre illuminator was used to illuminate the gratings at normal and oblique incidence. The angle of incidence was controlled by a Thorlabs FP90 adjustable flip platform, while the rotation angle of the sample was controlled by a Thorlabs RP005 manual rotation stage. The transmittance spectra of the color palette in Fig. 2 were measured with a 508 PV spectrophotometer from CRAIC Technologies.

**SEM imaging**

Before SEM imaging, the sample was sputtered with an Au:Pd 60:40 target (Ted Pella, Inc.) at 20 mA for 120s. The sample was then transferred to a JEOL JSM-7600F field emission scanning electron microscope system, which was then used to capture the SEM images in Figs. 3 – 4.

## AUTHOR INFORMATION

### Author Contributions

J.Y.E.C. did the experiments, analyzed the results, and wrote the manuscript. Q.F.R. assisted in the experiments, theory, and analysis of results. H.T.W., H.W., H.L.L. and Z.Y.Y. assisted in the analysis of results and drawing of schematics. Q.F.R, C.W.Q. and J.K.W.Y. supervised the research. All authors edited the manuscript.

### Notes

The authors declare no conflicts of interest.

## ACKNOWLEDGEMENTS

This research was supported by National Research Foundation (NRF) Singapore, under its Competitive Research Programme NRF-CRP001-021 and CRP20-2017-0004, as well as NRF Investigatorship Award NRF-NRFI06-2020-0005.

## REFERENCES

1. Fattal, D.; Peng, Z.; Tran, T.; Vo, S.; Fiorentino, M.; Brug, J.; Beausoleil, R. G. A multi-directional backlight for a wide-angle, glasses-free three-dimensional display. *Nature* **2013,** 495, (7441), 348-351.
2. Huang, Z.; Marks, D. L.; Smith, D. R. Out-of-plane computer-generated multicolor waveguide holography. *Optica* **2019,** 6, (2), 119-124.
3. Liu, Z.; Cui, Q.; Huang, Z.; Guo, L. J. Transparent Colored Display Enabled by Flat Glass Waveguide and Nanoimprinted Multilayer Gratings. *ACS Photonics* **2020,** 7, (6), 1418-1424.
4. Hua, J.; Hua, E.; Zhou, F.; Shi, J.; Wang, C.; Duan, H.; Hu, Y.; Qiao, W.; Chen, L. Foveated glasses-free 3D display with ultrawide field of view via a large-scale 2D-metagrating complex. *Light Sci. Appl.* **2021,** 10, (1), 213.

5.	Zeng, B.; Gao, Y.; Bartoli, F. J. Ultrathin Nanostructured Metals for Highly Transmissive Plasmonic Subtractive Color Filters. *Sci. Rep.* **2013,** 3, (1), 2840.
6.	Sturmberg, B. C. P.; Dossou, K. B.; Botten, L. C.; McPhedran, R. C.; de Sterke, C. M. Fano resonances of dielectric gratings: symmetries and broadband filtering. *Opt. Express* **2015,** 23, (24), A1672-A1686.
7.	Koirala, I.; Shrestha, V. R.; Park, C.-S.; Lee, S.-S.; Choi, D.-Y. Polarization-Controlled Broad Color Palette Based on an Ultrathin One-Dimensional Resonant Grating Structure. *Sci. Rep.* **2017,** 7, (1), 40073.
8.	Quaranta, G.; Basset, G.; Martin, O. J. F.; Gallinet, B. Color-Selective and Versatile Light Steering with up-Scalable Subwavelength Planar Optics. *ACS Photonics* **2017,** 4, (5), 1060-1066.
9.	Cao, X.; Du, Y.; Guo, Y.; Hu, G.; Zhang, M.; Wang, L.; Zhou, J.; Gao, Q.; Fischer, P.; Wang, J.; Stavrakis, S.; deMello, A. Replicating the Cynandra opis Butterfly's Structural Color for Bioinspired Bigrating Color Filters. *Adv. Mater.* **2022,** 34, (9), 2109161.
10.	Zhang, D.; Men, L.; Chen, Q. Femtosecond Laser Microfabricated Optofluidic Grating Refractometer. *IEEE Photon. Technol. Lett.* **2018,** 30, (4), 395-398.
11.	Zhang, D.; Men, L.; Chen, Q. Femtosecond laser fabricated polymeric grating for spectral tuning. *J. Phys. Commun.* **2018,** 2, (9), 095016.
12.	Quan, Y.-J.; Kim, Y.-G.; Kim, M.-S.; Min, S.-H.; Ahn, S.-H. Stretchable Biaxial and Shear Strain Sensors Using Diffractive Structural Colors. *ACS Nano* **2020,** 14, (5), 5392-5399.
13.	An, S.; Zhu, H.; Guo, C.; Fu, B.; Song, C.; Tao, P.; Shang, W.; Deng, T. Noncontact human-machine interaction based on hand-responsive infrared structural color. *Nat. Commun.* **2022,** 13, (1), 1446.
14.	Lu, Y. T.; Chi, S. Compact, reliable asymmetric optical configuration for cost-effective fabrication of multiplex dot matrix hologram in anti-counterfeiting applications. *Optik* **2003,** 114, (4), 161-167.
15.	Goncharsky, A.; Goncharsky, A.; Durlevich, S. Diffractive optical element with asymmetric microrelief for creating visual security features. *Opt. Express* **2015,** 23, (22), 29184-29192.
16.	Jiang, H.; Kaminska, B. Scalable Inkjet-Based Structural Color Printing by Molding Transparent Gratings on Multilayer Nanostructured Surfaces. *ACS Nano* **2018,** 12, (4), 3112-3125.
17.	Song, Q.; Pigeon, Y. E.; Heggarty, K. Faceted gratings for an optical security feature. *Appl. Opt.* **2020,** 59, (4), 910-917.
18.	Ushkov, A. A.; Verrier, I.; Kampfe, T.; Jourlin, Y. Subwavelength diffraction gratings with macroscopic moiré patterns generated via laser interference lithography. *Opt. Express* **2020,** 28, (11), 16453-16468.
19.	Rößler, F.; Kunze, T.; Lasagni, A. F. Fabrication of diffraction based security elements using direct laser interference patterning. *Opt. Express* **2017,** 25, (19), 22959-22970.
20.	Sharma, N.; Vangheluwe, M.; Vocanson, F.; Cazier, A.; Bugnet, M.; Reynaud, S.; Vermeulin, A.; Destouches, N. Laser-driven plasmonic gratings for hiding multiple images. *Mater. Horiz.* **2019,** 6, (5), 978-983.
21.	Tamulevičius, T.; Juodėnas, M.; Klinavičius, T.; Paulauskas, A.; Jankauskas, K.; Ostreika, A.; Žutautas, A.; Tamulevičius, S. Dot-Matrix Hologram Rendering Algorithm and its Validation through Direct Laser Interference Patterning. *Sci. Rep.* **2018,** 8, (1), 14245.
22.	Goncharsky, A.; Goncharsky, A.; Durlevich, S. Diffractive optical element for creating visual 3D images. *Opt. Express* **2016,** 24, (9), 9140-9148.
23.	Firsov, A.; Firsov, A.; Loechel, B.; Erko, A.; Svintsov, A.; Zaitsev, S. Fabrication of digital rainbow holograms and 3-D imaging using SEM based e-beam lithography. *Opt. Express* **2014,** 22, (23), 28756-28770.

24. Lu, W.-g.; Xiao, R.; Liu, J.; Wang, L.; Zhong, H.; Wang, Y. Large-area rainbow holographic diffraction gratings on a curved surface using transferred photopolymer films. *Opt. Lett.* **2018,** 43, (4), 675-678.
25. Jing, X.; Jin, Y. Transmittance analysis of diffraction phase grating. *Appl. Opt.* **2011,** 50, (9), C11-C18.
26. Meshalkin, A. Y.; Podlipnov, V. V.; Ustinov, A. V.; Achimova, E. A. Analysis of diffraction efficiency of phase gratings in dependence of duty cycle and depth. *J. Phys. Conf. Ser.* **2019,** 1368, (2), 022047.
27. Badloe, T.; Kim, J.; Kim, I.; Kim, W.-S.; Kim, W. S.; Kim, Y.-K.; Rho, J. Liquid crystal-powered Mie resonators for electrically tunable photorealistic color gradients and dark blacks. *Light Sci. Appl.* **2022,** 11, (1), 118.
28. Jung, Y.; Jung, H.; Choi, H.; Lee, H. Polarization Selective Color Filter Based on Plasmonic Nanograting Embedded Etalon Structures. *Nano Lett.* **2020,** 20, (9), 6344-6350.
29. Jung, C.; Yang, Y.; Jang, J.; Badloe, T.; Lee, T.; Mun, J.; Moon, S.-W.; Rho, J. Near-zero reflection of all-dielectric structural coloration enabling polarization-sensitive optical encryption with enhanced switchability. *Nanophotonics* **2021,** 10, (2), 919-926.
30. Ruan, Q.; Zhang, W.; Wang, H.; Chan, J. Y. E.; Wang, H.; Liu, H.; Fan, D.; Li, Y.; Qiu, C.-W.; Yang, J. K. W. Reconfiguring Colors of Single Relief Structures by Directional Stretching. *Adv. Mater.* **2022,** 34, (6), 2108128.
31. Fischer, J.; Wegener, M. Three-dimensional optical laser lithography beyond the diffraction limit. *Laser Photonics Rev.* **2013,** 7, (1), 22-44.
32. Liu, Y.; Lee, Y. H.; Lee, M. R.; Yang, Y.; Ling, X. Y. Flexible Three-Dimensional Anticounterfeiting Plasmonic Security Labels: Utilizing Z-Axis-Dependent SERS Readouts to Encode Multilayered Molecular Information. *ACS Photonics* **2017,** 4, (10), 2529-2536.
33. Lim, K. T. P.; Liu, H.; Liu, Y.; Yang, J. K. W. Holographic colour prints for enhanced optical security by combined phase and amplitude control. *Nat. Commun.* **2019,** 10, (1), 25.
34. Liu, Y.; Wang, H.; Ho, J.; Ng, R. C.; Ng, R. J. H.; Hall-Chen, V. H.; Koay, E. H. H.; Dong, Z.; Liu, H.; Qiu, C.-W.; Greer, J. R.; Yang, J. K. W. Structural color three-dimensional printing by shrinking photonic crystals. *Nat. Commun.* **2019,** 10, (1), 4340.
35. Chan, J. Y. E.; Ruan, Q.; Ng, R. J. H.; Qiu, C.-W.; Yang, J. K. W. Rotation-Selective Moiré Magnification of Structural Color Pattern Arrays. *ACS Nano* **2019,** 13, (12), 14138-14144.
36. Wang, H.; Wang, H.; Zhang, W.; Yang, J. K. W. Toward Near-Perfect Diffractive Optical Elements via Nanoscale 3D Printing. *ACS Nano* **2020,** 14, (8), 10452-10461.
37. Zhang, Y.; Jiao, Y.; Li, C.; Chen, C.; Li, J.; Hu, Y.; Wu, D.; Chu, J. Bioinspired micro/nanostructured surfaces prepared by femtosecond laser direct writing for multi-functional applications. *Int. J. Extrem. Manuf.* **2020,** 2, (3), 032002.
38. Chan, J. Y. E.; Ruan, Q.; Jiang, M.; Wang, H.; Wang, H.; Zhang, W.; Qiu, C.-W.; Yang, J. K. W. High-resolution light field prints by nanoscale 3D printing. *Nat. Commun.* **2021,** 12, (1), 3728.
39. Lao, Z.; Sun, R.; Jin, D.; Ren, Z.; Xin, C.; Zhang, Y.; Jiang, S.; Zhang, Y.; Zhang, L. Encryption/decryption and microtarget capturing by pH-driven Janus microstructures fabricated by the same femtosecond laser printing parameters. *Int. J. Extrem. Manuf.* **2021,** 3, (2), 025001.
40. Wang, H.; Ruan, Q.; Wang, H.; Rezaei, S. D.; Lim, K. T. P.; Liu, H.; Zhang, W.; Trisno, J.; Chan, J. Y. E.; Yang, J. K. W. Full Color and Grayscale Painting with 3D Printed Low-Index Nanopillars. *Nano Lett.* **2021,** 21, (11), 4721-4729.
41. Wang, H.; Wang, H.; Ruan, Q.; Tan, Y. S.; Qiu, C.-W.; Yang, J. K. W. Optical Fireworks Based on Multifocal Three-Dimensional Color Prints. *ACS Nano* **2021,** 15, (6), 10185-10193.
42. Zhang, W.; Wang, H.; Wang, H.; Chan, J. Y. E.; Liu, H.; Zhang, B.; Zhang, Y.-F.; Agarwal, K.; Yang, X.; Ranganath, A. S.; Low, H. Y.; Ge, Q.; Yang, J. K. W. Structural multi-colour invisible inks with submicron 4D printing of shape memory polymers. *Nat. Commun.* **2021,** 12, (1), 112.

43. Jung, C.; Kim, G.; Jeong, M.; Jang, J.; Dong, Z.; Badloe, T.; Yang, J. K. W.; Rho, J. Metasurface-Driven Optically Variable Devices. *Chem. Rev.* **2021,** 121, (21), 13013-13050.
44. Jang, J.; Jeong, H.; Hu, G.; Qiu, C.-W.; Nam, K. T.; Rho, J. Kerker-Conditioned Dynamic Cryptographic Nanoprints. *Adv. Opt. Mater.* **2019,** 7, (4), 1801070.
45. Harvey, J.; Pfisterer, R. Understanding diffraction grating behavior: including conical diffraction and Rayleigh anomalies from transmission gratings. *Opt. Eng.* **2019,** 58, (8), 087105.
46. Knop, K. Rigorous diffraction theory for transmission phase gratings with deep rectangular grooves. *J. Opt. Soc. Am.* **1978,** 68, (9), 1206-1210.
47. Pommet, D. A.; Moharam, M. G.; Grann, E. B. Limits of scalar diffraction theory for diffractive phase elements. *J. Opt. Soc. Am. A* **1994,** 11, (6), 1827-1834.
48. Harvey, J. E.; Krywonos, A.; Bogunovic, D. Nonparaxial scalar treatment of sinusoidal phase gratings. *J. Opt. Soc. Am. A* **2006,** 23, (4), 858-865.
49. Ruan, D.; Zhu, L.; Jing, X.; Tian, Y.; Wang, L.; Jin, S. Validity of scalar diffraction theory and effective medium theory for analysis of a blazed grating microstructure at oblique incidence. *Appl. Opt.* **2014,** 53, (11), 2357-2365.
50. Painting of elephant inside snake. Free image on Shutterstock - 109274537. https://www.shutterstock.com/image-vector/snake-swallow-elephant-109274537
51. Dehaeck, S.; Scheid, B.; Lambert, P. Adaptive stitching for meso-scale printing with two-photon lithography. *Addit. Manuf.* **2018,** 21, 589-597.
52. Painting of giraffe. Free image on Peakpx - 530915. https://www.peakpx.com/530915/giraffe-head-painting
53. Painting of snake. Free image on Pixabay - 6095. https://pixabay.com/photos/snake-cartoon-character-drawing-6095/
54. Painting of turtle. Free image on Pixabay - 6099. https://pixabay.com/photos/turtle-cartoon-character-drawing-6099/
55. Painting of sheep. Free image on Pixabay - 6101. https://pixabay.com/photos/sheep-cartoon-character-drawing-6101/
56. Image of Rubin's vase. Free image on Wikimedia Commons - Facevase. https://commons.wikimedia.org/wiki/File:Facevase.JPG
57. Geng, Q.; Wang, D.; Chen, P.; Chen, S.-C. Ultrafast multi-focus 3-D nano-fabrication based on two-photon polymerization. *Nat. Commun.* **2019,** 10, (1), 2179.
58. Saha, S. K.; Wang, D.; Nguyen, V. H.; Chang, Y.; Oakdale, J. S.; Chen, S.-C. Scalable submicrometer additive manufacturing. *Science* **2019,** 366, (6461), 105-109.
59. Zhang, L.; Liu, B.; Wang, C.; Xin, C.; Li, R.; Wang, D.; Xu, L.; Fan, S.; Zhang, J.; Zhang, C.; Hu, Y.; Li, J.; Wu, D.; Zhang, L.; Chu, J. Functional Shape-Morphing Microarchitectures Fabricated by Dynamic Holographically Shifted Femtosecond Multifoci. *Nano Lett.* **2022,** 22, (13), 5277-5286.

# Supporting Information:
# Full Geometric Control of Hidden Color Information in Diffraction Gratings under Angled White Light Illumination

**Authors:** John You En Chan [1], Qifeng Ruan [1,2]*, Hongtao Wang [1,3], Hao Wang [1], Hailong Liu [4], Zhiyuan Yan [3], Cheng-Wei Qiu [3], Joel K.W. Yang [1,4]*

[1] Engineering Product Development, Singapore University of Technology and Design, Singapore 487372.

[2] Ministry of Industry and Information Technology Key Lab of Micro-Nano Optoelectronic Information System, Harbin Institute of Technology (Shenzhen), Shenzhen 518055, P. R. China.

[3] Department of Electrical and Computer Engineering, National University of Singapore, Singapore 117583.

[4] Institute of Materials Research and Engineering, A*STAR (Agency for Science, Technology and Research), Singapore 138634.

* To whom correspondence should be addressed. Email: ruanqifeng@hit.edu.cn; joel_yang@sutd.edu.sg

## Derivation of Equation 1

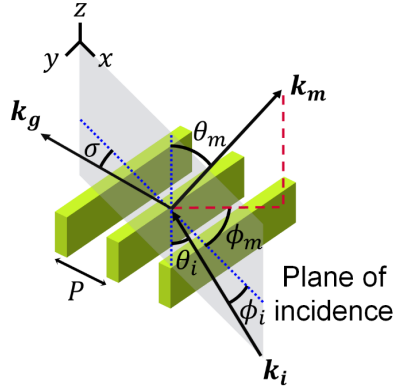

The momentum matching condition for a set of diffraction gratings is expressed in Equation S1:

$$\bm{k_m} = \bm{k_i} - m\bm{k_g} \tag{S1}$$

where $\bm{k_m}$ and $\bm{k_i}$ are the wave vectors of the $m^{th}$ order diffracted light and incident light respectively; $\bm{k_g}$ is the grating vector. Expressing each vector with magnitude and direction yields Equation S2:

$$n_e \frac{2\pi}{\lambda} \begin{pmatrix} \sin\theta_m \cos\phi_m \\ \sin\theta_m \sin\phi_m \\ \cos\theta_m \end{pmatrix} = n_e \frac{2\pi}{\lambda} \begin{pmatrix} \sin\theta_i \cos\phi_i \\ \sin\theta_i \sin\phi_i \\ \cos\theta_i \end{pmatrix} - m \frac{2\pi}{P} \begin{pmatrix} \cos\sigma \\ \sin\sigma \\ 0 \end{pmatrix} \tag{S2}$$

where $\theta_{i(m)}$ and $\phi_{i(m)}$ are the polar and azimuthal angles of incident (diffracted) light respectively; $\lambda$ is the wavelength of light; $P$ is the periodicity of the grating. The environment is assumed to be air (refractive index $n_e = 1$).

## Derivation of Equation 4

Assume that normally incident light ($\theta_i = 0°$) illuminates a binary phase grating (refractive index $n_g$) in air (refractive index $n_e$). The grating has height $H$ and periodicity $P$. Light that passes through the grating interferes with light that passes through air. Constructive interference occurs when the phase difference between the two light paths is equal to an integer multiple of $2\pi$. This condition is expressed in Equation S3:

$$\Delta\rho = \frac{2\pi}{\lambda}(n_g - n_e)H = q2\pi \tag{S3}$$

where $q$ is a positive integer and $\lambda$ is the wavelength of light. By rearranging Equation S3, the wavelength at which constructive interference occurs is expressed in Equation S4:

$$\lambda = \lambda_0 = \frac{H(n_g - n_e)}{q} \tag{S4}$$

## Derivation of Equation 6

When the grating vector is parallel to the plane of incidence ($\sigma = 0°$), diffraction simplifies to the classical case which is expressed in Equation S5:

$$\sin\theta_i - \sin\theta_m = -\frac{m\lambda}{P} \tag{S5}$$

where $\theta_i$ is the angle of incidence, $\theta_m$ is the angle of the $m^{th}$ order diffracted light, $P$ is the periodicity of the grating, and $\lambda$ is the wavelength of light. Only wavelengths of diffracted light that are collected by the microscope objective contribute to the observed colour of the grating. The range of angles for diffracted light collected by the microscope objective is determined by its numerical aperture ($NA$) in air, which is expressed in Equation S6:

$$NA = |\sin\theta_m| \tag{S6}$$

By substituting Equation S6 into Equation S5, the upper and lower limits of collected wavelengths are expressed in Equations S7 – S8 respectively.

$$\lambda_{m,u} = \frac{P(\sin\theta_i + NA)}{m} \tag{S7}$$

$$\lambda_{m,l} = \frac{P(\sin\theta_i - NA)}{m} \tag{S8}$$

Hence, the collected range of the $m^{th}$ order diffracted wavelengths is expressed in Equation S9:

$$\Delta\lambda_m = \lambda_{m,u} - \lambda_{m,l} = \frac{2P(NA)}{m} \tag{S9}$$

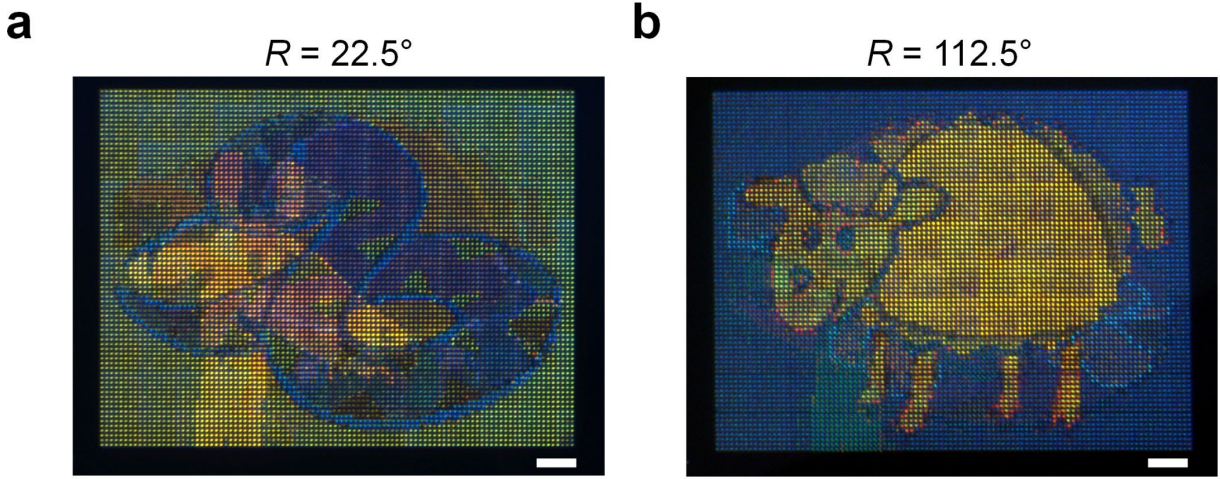

**Fig. S1. Optical micrographs of image crosstalk at sample rotation angle *R* in between adjacent grating sectors.** (**a**) *R* = 22.5°. Scale bar: 100 μm. (**b**) *R* = 112.5°. Scale bar: 100 μm.

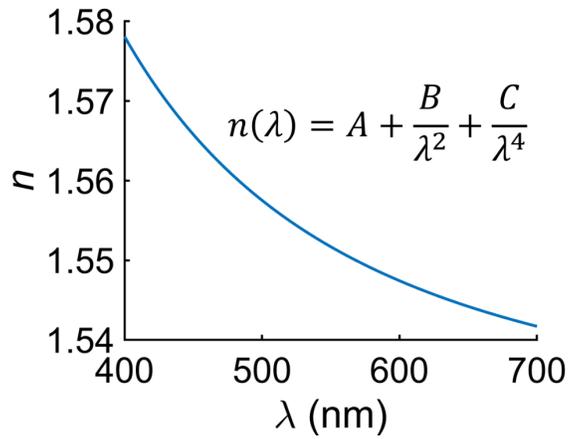

$$n(\lambda) = A + \frac{B}{\lambda^2} + \frac{C}{\lambda^4}$$

**Fig. S2. Refractive index of UV cured IP-Dip photoresist.** The Cauchy parameters are given by *A* = 1.5273, *B* = 6.5456*10$^{-3}$ μm$^2$, *C* = 2.5345*10$^{-4}$ μm$^4$ from Ref.[1].

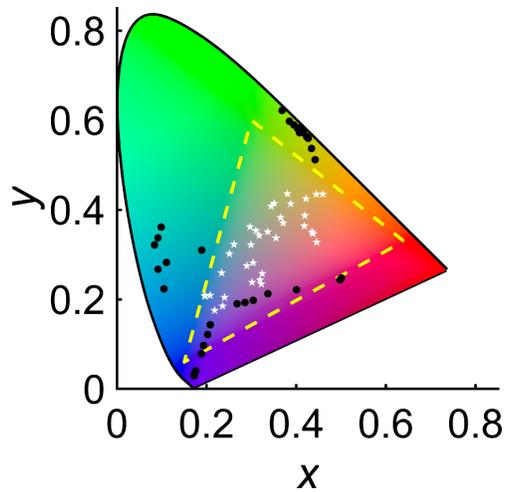

**Fig. S3. CIE 1931 chromaticity diagram.** The chromaticity coordinates of the color palettes under normal ($\theta_i = 0°$, white stars) and oblique incidence ($\theta_i = 45°$, black dots) are plotted for CIE D65 illuminant and CIE 2° standard observer. The sRGB color space is marked by the yellow dashed outline of a triangle.

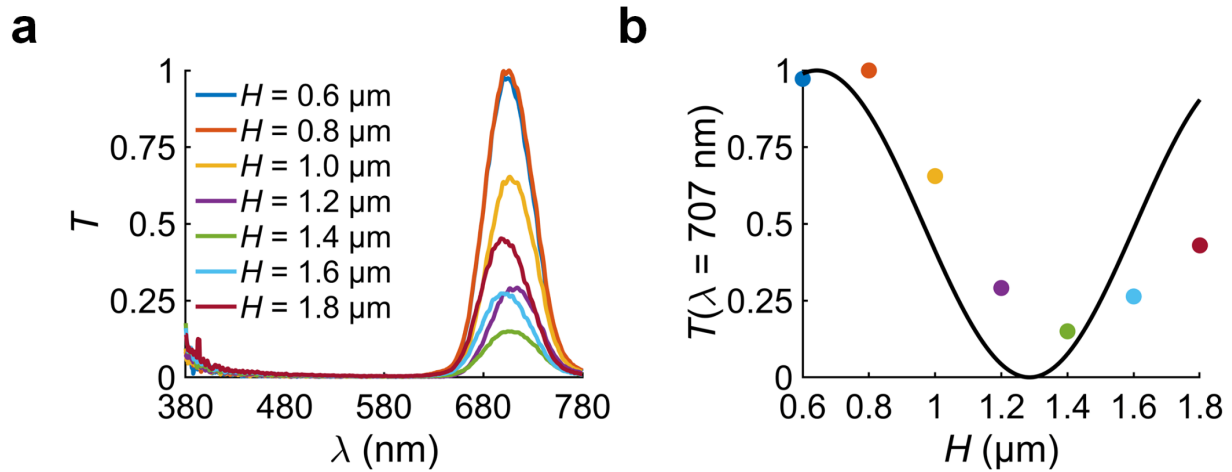

**Fig. S4. Effect of varying grating height under oblique incidence.** The grating height $H = 0.6 - 1.8$ µm, grating periodicity = 1.0 µm, angle of incidence = 45°. (**a**) Normalized transmittance spectra of the gratings. (**b**) Graph of normalized transmittance $T(\lambda = 707$ nm$)$ vs. $H$. The black curve is the approximate intensity distribution calculated from scalar diffraction theory.

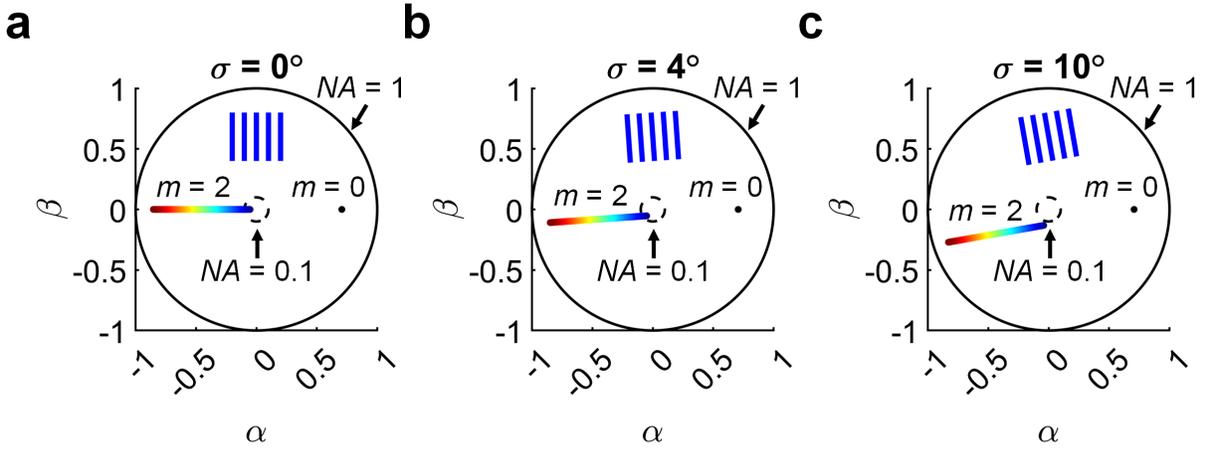

**Fig. S5. Directional cosine diagrams for the selected grating (periodicity $P$ = 1.0 μm) with varying $\sigma$ (the azimuthal angle between the grating vector and plane of incidence).** (**a**) $\sigma$ = 0° (**b**) $\sigma$ = 4° (**c**) $\sigma$ = 10°. The diagrams are plotted for oblique incidence $\theta_i$ = 45°, wavelengths $\lambda$ = 380 – 780 nm, diffraction orders $m$ = [0, 2]. In each diagram, the smaller dashed circle represents numerical aperture $NA$ = 0.1, and the larger solid circle $NA$ = 1. Only spectra lines that lie inside the $NA$ = 0.1 region are collected by the microscope objective and contribute to the observed color. Spectra lines that lie outside the $NA$ = 1 region are evanescent.

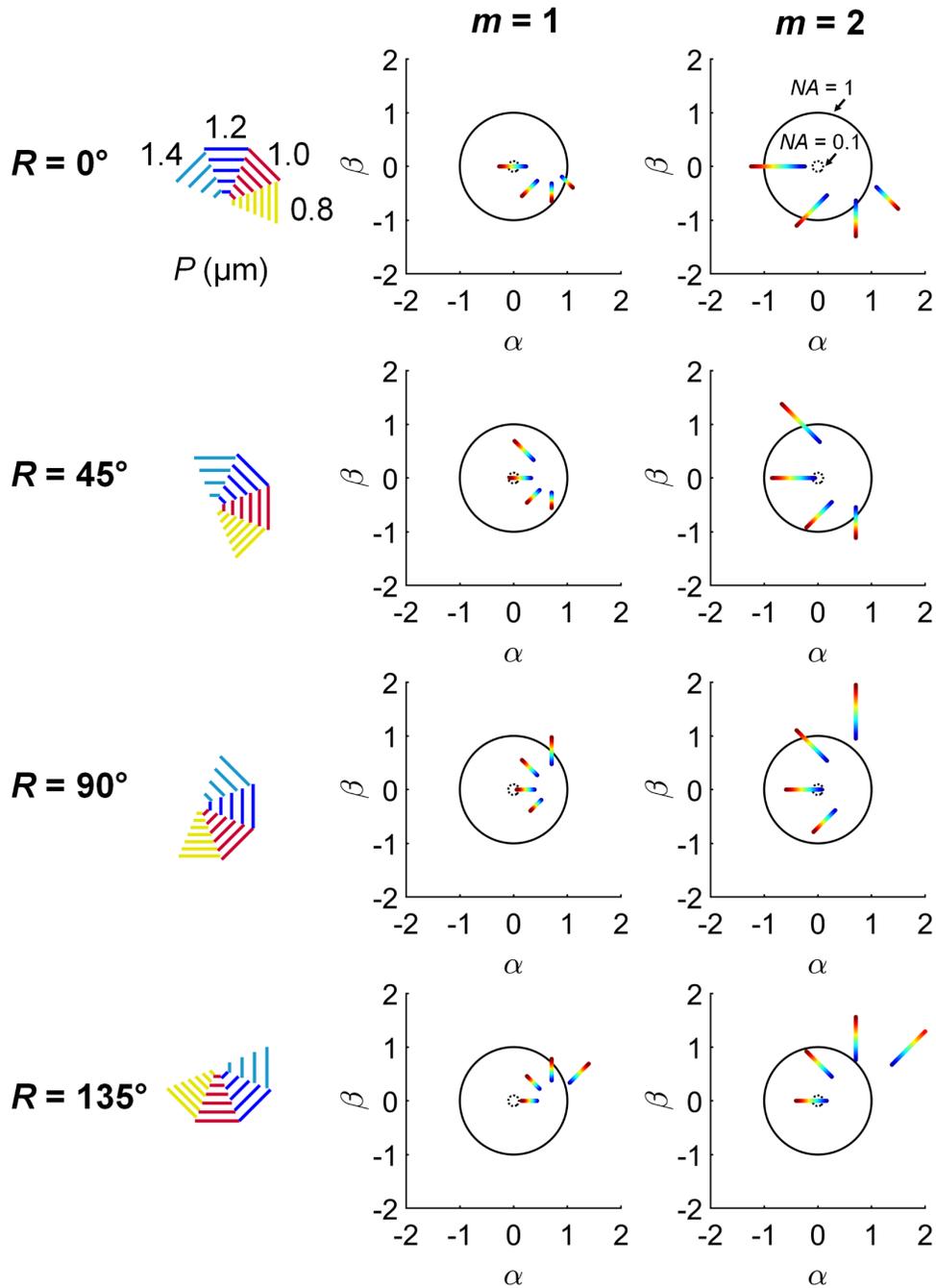

**Fig. S6. Directional cosine diagrams for a pixel comprising four angularly multiplexed gratings with periodicities $P$ = [0.8, 1.0, 1.2, 1.4] µm.** The diagrams are plotted for oblique incidence $\theta_i$ = 45°, wavelengths $\lambda$ = 380 – 780 nm, diffraction orders $m$ = [1, 2], pixel rotation angles $R$ = [0, 45, 90, 135] °. In each diagram, the smaller dotted circle represents numerical aperture $NA$ = 0.1, and the larger solid circle $NA$ = 1. Only spectra lines that lie inside the $NA$ = 0.1 region are collected by the microscope objective and contribute to the observed color. Spectra lines that lie outside the $NA$ = 1 region are evanescent.

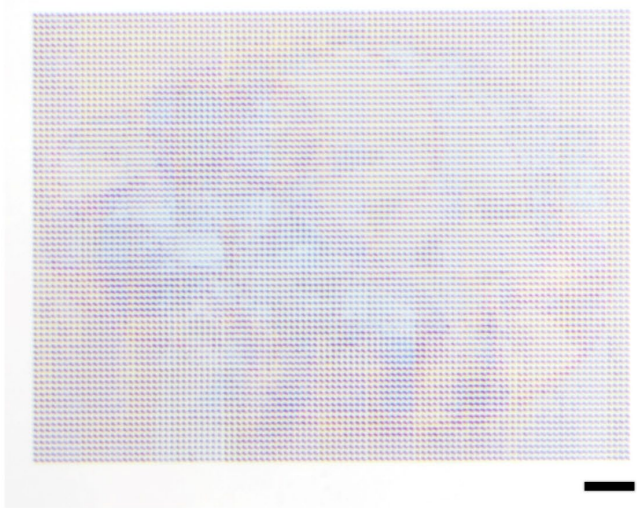 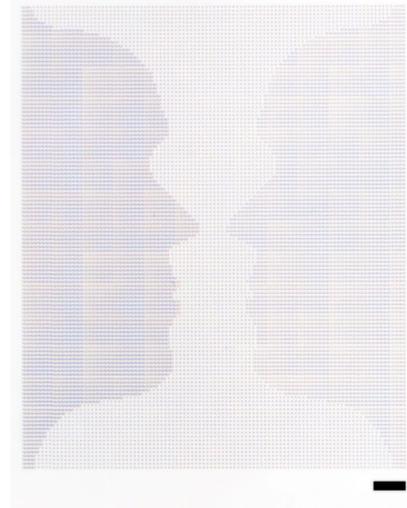

**Fig. S7. Optical micrographs of samples with angularly multiplexed gratings under normal incidence.** (**a**) Image encoded with animal paintings (giraffe, snake, turtle, sheep). Scale bar: 100 µm. (**b**) Image of Rubin's vase. Scale bar: 100 µm.

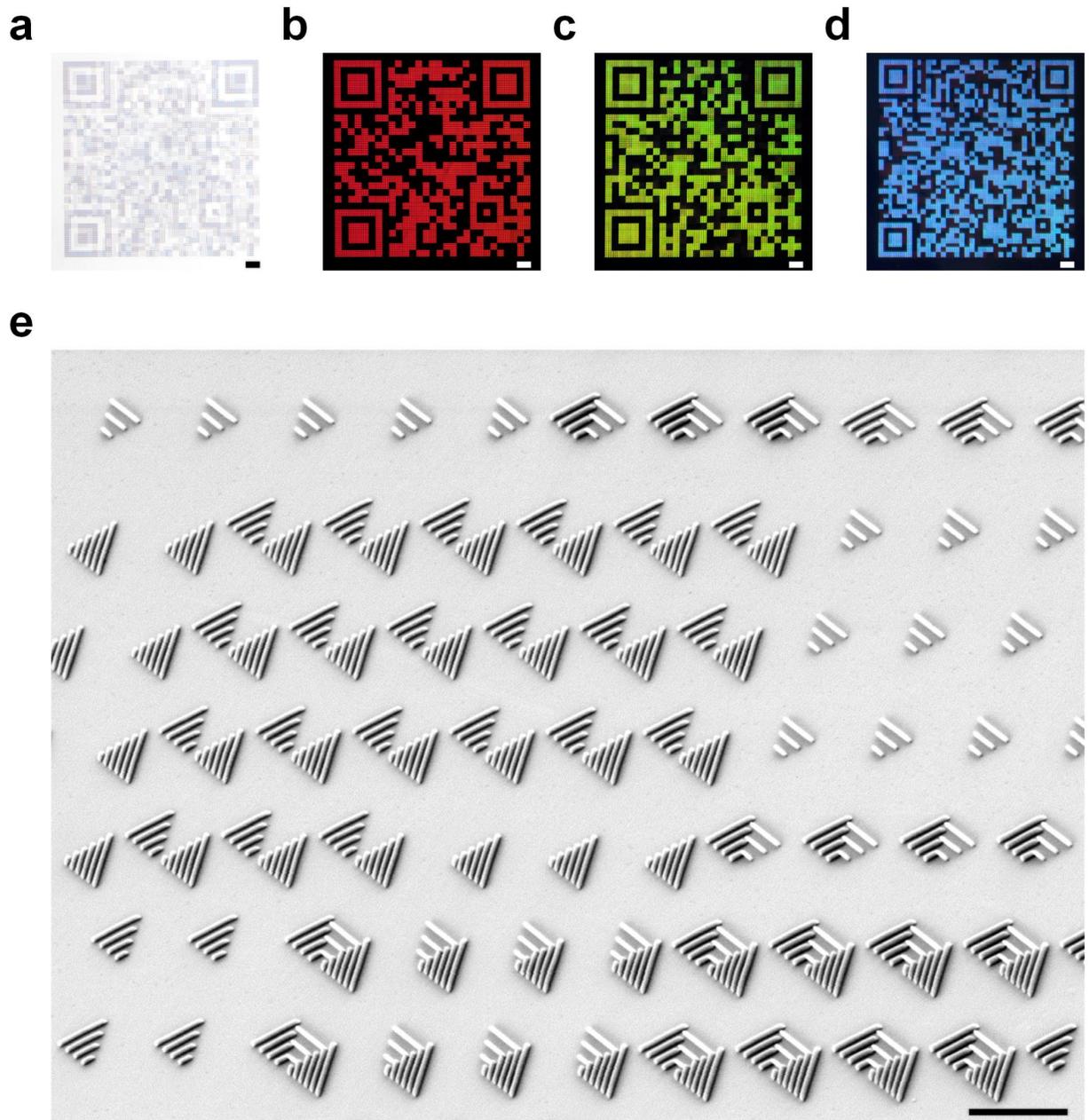

**Fig. S8. Quick response codes encoded with angularly multiplexed gratings.** (**a**) Optical micrograph of the quick response code under normal incidence ($\theta_i = 0°$). Scale bar: 100 μm. (**b – d**) Optical micrograph of the quick response code under oblique incidence ($\theta_i = 45°$) by rotating the sample in angular steps of 60°. Scale bar: 100 μm. (**e**) SEM images of gratings in the sample. Scale bar: 10 μm.

**References**

1. Gissibl, T.; Wagner, S.; Sykora, J.; Schmid, M.; Giessen, H. Refractive index measurements of photo-resists for three-dimensional direct laser writing. *Opt. Mater. Express* **2017,** 7, (7), 2293-2298.